\documentstyle[preprint,aps]{revtex}
\begin{document}
\title{Nucleation at Finite Temperature Beyond the Superminispace Model}
\author{J.--Q. Liang $^{1,2}$\footnote{email: jqliang@mail.sxu.edu.cn},
 H. J. W. M\"{u}ller--Kirsten $^{1}
$\footnote{email: mueller1@physik.uni--kl.de}, D. K. Park
$^{1,3}$\footnote{email: dkpark@genphys.kyungnam.ac.kr} and
 A. V. Shurgaia $^{1}$\footnote{email: shurgaia@physik.uni--kl.de}}
\address{\footnotesize{1. Department of Physics,
 University of Kaiserslautern, D--67653 Kaiserslautern, Germany\\
2. Department of Physics, and Institute of Theoretical Physics, Shanxi University, Taiyuan, Shanxi, 030006, P.R. China\\
3. Department of Physics, Kyungnam University, Masan, 631--701, Korea}}
\maketitle
\begin{abstract}
The transition from the quantum to the classical
 regime of the nucleation  of the closed Robertson--Walker
 Universe with spacially homogeneous matter fields is investigated
with a perturbation expansion around the sphaleron
 configuration. A criterion is derived for the occurrence of
a first--order type transition, and the related phase diagram for
  scalar and vector 
fields is obtained. For scalar fields both the first and second order
 transitions can occur 
 depending on the shape of the potential barrier. For a vector
 field, here that of an $O(3)$ nonlinear 
$\sigma$--model, the transition is seen to be only of the first order.
\newline
PACS numbers: 11.15.Kc, 03.65Sq, 05.70.Fh, 98.80.Cq
\end{abstract}

\newpage
\section{Introduction}
\label{sec: 1}
The problem of the decay rate of a metastable state
  and coherence of degenerate states via quantum
 tunneling has profound physical implications in many fundamental
phenomena in various branches of physics 
as e.g. in condensed matter and particle physics and cosmology. 
The instanton techniques initiated long ago by 
Langer\cite{langer} and 
Coleman\cite{coleman} are a major
tool of investigation,
and provide us with a formalism capable
 of producing accurate values for the tunneling rate.
 The instanton method has been widely used to study tunneling at zero
temperature. The generalization to
 finite temperature tunneling has been
 a long--standing problem in which a new type of solution
 satisfying a periodic boundary
condition, and therefore called the periodic instanton,
 was gradually realized to be relevant\cite{manton,yu}. 
The exact analytic form of the periodic instanton is known 
only in one--dimensional
quantum mechanics\cite{liang1}. In field theory models,
 it can be found either approximately
 at low energies or numerically. Thus quantum tunneling
 at finite temperature 
$T$ is, under certain conditions, 
dominated by periodic instantons with finite energy $E$, and
 in the semi--classical approximation the euclidean action
 is expected to be saturated by a single periodic
instanton. Thus only periodic instantons with the period equal
 to the inverse temperature can dominate
 the thermal rate. With exponential accuracy the tunneling probability 
$P(E)$ at a given energy $E$ can be written as
\begin{equation}
\label{1}
P(E)\sim e^{-W(E)}=e^{-S(\beta)+E\beta}
\end{equation}
The period $\beta $ of the periodic instanton is
 related to the energy $E$ in the standard
 way $E=\frac{\partial S}{\partial \beta}$ and
 $S(\beta )$ is the action of the periodic
instanton per period. With increasing temperature
 thermal hopping becomes more and more important
 and beyond some critical or crossover temperature $T_{c}$ becomes the
decisive mechanism. The barrier penetration process
 is then governed by a static solution of the euclidean field
 equation, i.e. the sphaleron. The study of the crossover
from quantum tunneling to thermal hopping is an
 interesting subject with a long history\cite{affle,linde}.
 Under certain assumptions for the shape of 
the potential barrier, it
was found that the transition between quantum tunneling
 and thermally assisted hopping occurs at the temperature $T_{c}$ 
and was recognized as a smooth second order
transition in the quantum mechanical models of Affleck\cite{affle} 
and the cosmological models of Linde\cite{linde}.
 It was demonstrated that the periodic instantons which
 govern the tunneling in the
intermediate temperature regime interpolate smoothly
 between the zero temperature or vacuum 
instanton and the sphaleron.
 In analogy with the terminology of statistical mechanics this
phenomenon can be referred to as a second--order
 transition characterized by the plot of euclidean action
 $S(\beta )$ versus instanton period $\beta$, the latter being the inverse
temperature in the finite temperature field theory.\\
 However, it was shown \cite{chudn1} that the smooth
 transition is not generic. 
Using a simple quantum mechanical model it was demonstrated
 that the time derivative 
of the euclidean
action would be discontinuous if the period of the instanton
 is not a monotonic function of energy. Assuming
 that there exists a minimum of $\beta(E)$, i. e.
that $\frac{d\beta}{dE}=0$ at some value of $E$, the
second time derivative of the action 
\begin{equation}
\label{2}
\frac{d^{2}S(\beta)}{d\beta^{2}}=\frac{1}{\frac{d\beta}{dE}}
\end{equation}
would not be defined at the minimum, or, in other words,
 the first time derivative is discontinuous.
 The sharp first order transition occurs as a bifurcation in the plot of
the action $S(\beta)$ versus the period  $\beta$. In the
 context of field theory the crossover behaviour
 and the bifurcation of periodic instantons have also been explained in a
more transparent manner\cite{kuzne}. The idea to determine 
the order of a transition from the plot of euclidean
 action versus the period of the instantons was subsequently extended,
 and a
sufficient condition for the existence of a
 first--order transition was derived using only small fluctuations
 around the sphaleron. If the period $\beta (E\rightarrow U_{0})$ of the
periodic instanton close to the barrier peak can be found, a
 sufficient condition to have a first--order transition
 is seen to be that $\beta (E\rightarrow U_{0})-\beta_{s}<0$
or $\omega^{2}-\omega^{2}_{s} >0$, where
 $U_{0}$ denotes the barrier height and 
$\beta_{s}$ is the period of small oscillation around
 the sphaleron\cite{gorok,muller};
 $\omega$ and $\omega_{s}$
are the corresponding frequencies. This observation
 triggered  active research on the
transition behaviour,
as e.g. in connection with spin tunneling in condensed matter
 physics\cite{chudn2,liang2,lee1} and with tunneling in various field theory
models\cite{lee2,park1,park2,garri,ferre}.\\
It is also interesting to investigate the crossover from
 quantum tunneling to thermal hopping in the context of cosmology.
 After the pioneering attempt of Linde\cite{linde} little
work has been done along this direction in view
 of the lack of solvable models. Motivated by
a similar study of bubble nucleation in field theory\cite{garri,ferre},
 the nucleation rate in
a superminispace model\cite{vilen} has been extended
 to finite temperature\cite{liang3} where the matter field
 is frozen out and leaves only  the constant vacuum energy density.
 In this context a
periodic instanton solution was obtained analytically,
 and the transition from tunneling to thermal hopping
 was found to be of the first order.
 The  superminispace model may be
 too simple to
 imply a realistic result. In the present paper we
therefore extend the model to one including a 
spacially homogeneous matter field. Small fluctuations
 around the sphaleron solutions
are then studied with a perturbation method. Criteria
 for a first--order transition and related phase diagrams
 are obtained for both scalar and vector fields. This
investigation may shed light on our understanding
 of the time evolution of the early Universe. In Sec. 2 
the effective Lagrangian and the equations of motion are obtained for the 
closed Robertson--Walker (RW) metric interacting with spacially
 homogeneous scalar and vector fields. The physical meaning
 of tunneling is explained. The oscillation frequencies
around the sphaleron are obtained with a perturbation expansion 
in Sec. 3. We derive the criterion for the first--order
  transition of the nucleation and the related phase diagram
 for interaction with 
a scalar field.
 In Sec. 4 we apply a similar approach to the model with a
vector field.
\section{Sphalerons and the Thermal Rate of Nucleation}
\label{sec:2}
We begin with the model of the Universe defined by the action,
\begin{equation}
\label{3}
S=\int d^{4}x\sqrt {-g}\left [ -\frac {\cal R}{16\pi G}+{\cal L}_{m}\right ]
\end{equation}
where $\cal {R}$ is the Ricci scalar. The Lagrangian
 density of the  scalar matter field $\phi$ is of the general form,
\begin{equation}
\label{4}
{\cal L}_{m}=\frac{1}{G_{\phi}}\left[\frac{1}{2}\partial_{\mu}\phi \partial^{\mu}\phi -V(\phi )\right]
\end{equation}
where $G_{\phi}$ is a dimensional parameter.
 For a vector field we consider that of the $O(3)$ nonlinear $\sigma$--model
 with a symmetry breaking term such as,
\begin{equation}
\label{5}
{\cal L}_{m}=\frac{1}{2}m\sum_{a}\partial_{\mu}n_{a}\partial^{\mu}n_{a}
-\frac{1}{\lambda^{2}}(1+n_{3}),\;\;\;
\sum^{3}_{a=1}n^{2}_{a}=1
\end{equation}
where m and $\lambda$ are suitable parameters.
 Contemporary cosmological models are based
 on the idea that the Universe is pretty
 much the same everywhere. More
mathematically precise properties
 of the manifold may be isotropy
 and homogeneity. The spacetime to be considered here is
 ${\bf R}\times \Sigma$ where ${\bf R}$
represents the time direction and $\Sigma$ is a homogeneous and
 isotropic 3--manifold. The Universe is also assumed
 to be closed. We therefore obtain the
Robertson--Walker (RW) metric of the closed case,
\begin{equation}
\label{6}
ds^{2}=dt^{2}-R^{2}(t)d\Omega^{2}_{3}
\end{equation}
The manifold $\Sigma$ in our case is a three--sphere
 $S^{3}$ and the lapse function is simply equal to 1.  R(t) is known as the scale 
factor which tells us ``how big'' the spacetime slice $\Sigma $ is at time t. 
$d\Omega^{2}_{3}$ is the metric on a unit 3-sphere. The
 Ricci scalar is found to be
\begin{equation}
\label{7}
{\cal R}=-6\left [ \frac{\ddot{R}}{R}+
\frac{\dot{R}^{2}}{R^{2}}+\frac{1}{R^{2}}\right ]
\end{equation}
where a dot denotes the
 time derivative.  For spacially homogeneous matter
 fields $\phi=\phi (t)$ and ${\bf n}={\bf n}(t)$ the angle
 integrals can be carried out and we
have,
\begin{equation}
\label{8}
S=\int Ldt
\end{equation}
The effective Lagrangians are obtained as
\begin{equation}
\label{9}
L=2\pi^{2}\left \{ -\frac{3R(\dot{R}^{2}-1)}{8\pi G}+\frac {1}{G_{\phi}}\left [\frac{1}{2}R^{3}\dot{\phi}^{2}-R^{3}V(\phi)\right ]\right \}
\end{equation}
for the scalar field and
\begin{equation}
\label{10}
L=2\pi^{2}\left [ -\frac{3R(\dot{R}^{2}-1)}{8\pi G}+\frac{mR^{3}}{2}\sum_{a}\dot{n}^{2}_{a}-\frac{R^{3}}{\lambda^{2}}(1+n_{3})\right ]
\end{equation}
for the vector field. The canonical momenta are defined by
\begin{equation}
\label{11}
P_{R}=\frac{\partial L}{\partial \dot{R}}
=-\frac{3\pi R\dot{R}}{2G},\hspace{0.2cm} P_{\phi}=\frac {\partial L}{\partial
\dot{\phi}}=\frac{2\pi^{2}}{G_{\phi}}R^{3}\dot{\phi},\hspace{0.2cm}
P_{a}=2\pi^{2}mR^{3}\dot{n}_{a}
\end{equation}
The corresponding Hamiltonians
\begin{equation}
\label{12}
H=\frac{G_{\phi}}{4\pi^{2}R^{3}}P^{2}_{\phi}
-\frac{G}{3\pi R}P^{2}_{R}-\frac{3\pi}{4G}R
+\frac{2\pi^{2}}{G_{\phi}}R^{3}V(\phi),
\end{equation}
\begin{equation}
\label{13}
H=\frac{1}{4\pi^{2}mR^{3}}P^{2}_{a}
-\frac{G}{3\pi R}P^{2}_{R}-\frac{3\pi}{4G}R
+\frac{2\pi^{2}}{\lambda^{2}}R^{3}(1+n_{3})
\end{equation}
are conserved quantities. For our purposes of the
study of nucleation we make use of the Wick
 rotation $\tau =it$ and obtain the euclidean Lagrangians,
\begin{equation}
\label{14}
L_{e}=2\pi^{2}\left \{-\frac{3R(\dot{R}^{2}+1)}{8\pi G}
+\frac{1}{G_{\phi}}\left[\frac{1}{2}R^{3}\dot{\phi}^{2}
+R^{3}V(\phi)\right ]\right \},
\end{equation}
\begin{equation}
\label{15}
L_{e}=2\pi^{2}\left [-\frac{3R(\dot{R}^{2}+1)}{8\pi G}
+\frac{m}{2}R^{3}\sum_{a}\dot{n}^{2}_{a}
+\frac {R^{3}}{\lambda^{2}}(1+n_{3})\right ]
\end{equation}
From now on the dot denotes imaginary time derivatives, e. g.
 $\dot{R}=\frac {dR}{d\tau}$.\\
The euclidean equations of motion for the
 scalar field are seen to be
\begin{equation}
\label{16}
\frac {d}{d\tau}(R\dot{R})
-\frac {\dot{R}^{2}+1}{2}+2\pi \tilde{G}R^{3}\dot{\phi}^{2}
+4\pi \tilde{G}R^{2}V(\phi)=0
\end{equation}
where $\tilde{G}=\frac{G}{G_{\phi}}$, and
\begin{equation}
\label{17}
\frac {d}{d\tau}(R^{3}\dot{\phi})=R^{3}\frac {\partial V}{\partial \phi}
\end{equation}
The sphalerons $\phi_{0}$ and $R_{0}$ are static
 solutions of eqs. (16) and (17)
 with $\dot{\phi}=\ddot{\phi}=\dot{R}=\ddot{R}=0$. From
 eq.(16) we have
\begin{equation}
\label{18}
R_{0}=\left [\frac {1}{8\pi \tilde{G}V(\phi_{0})}\right ]^{\frac {1}{2}}
\end{equation}
$\phi_{0}$ is the position of the peak of the
 potential barrier such that
 $\frac {\partial V}{\partial \phi}\vert _{\phi=\phi _{0}}= 0$. With the sphaleron $\phi_{0}$ the
effective potential of the dynamical
 variable $R$ is seen to be from eq.(9),
 \begin{equation}
\label{19}
U(R)=-\frac{R^{3}V(\phi_{0})}{G_{\phi}}+\frac{3R}{8\pi G}
\end{equation}
$R_{0}$ is just the position of the above
 potential barrier peak and indeed the sphaleron.
 The thermal rate of nucleation at temperature T is given by
the  Arrhenius law,
\begin{equation}
\label{20}
P(T)\sim e^{-\frac {U(R_{0})}{T}}
\end{equation}
Our superminispace model here is simply
 the dynamical model described by the equation of motion (16)
 with the scalar field variable $\phi$ in $V(\phi)$ replaced by the sphaleron
 $\phi_{0}$. The
nucleation process in the 
superminispace model has been extended to the finite
 temperature case with the periodic instanton formalism
 in our previous work\cite{liang3}. 
In the present paper the scalar field is not
frozen out and we instead investigate the fluctuation of the fields
 around the sphalerons. The crossover behaviour from
 tunneling to thermal hopping can be obtained with
perturbation expansions.

\section{Nucleation at finite temperature in presence of a scalar field}
\label{sec:3}
As we demonstrated above, the crossover behaviour
 of the nucleation of our model Universe from
 quantum tunneling to thermal activation can be obtained from the
deviation of the period of the periodic instanton
from that of the sphaleron. To this end we expand the field
 variables about the sphaleron configurations
$\phi_{0}$ and $R_{0}$, i. e. we set
\begin{equation}
\label{21}
\phi=\phi_{0}+\xi, \hspace{1cm} R=R_{0}+\eta
\end{equation}
where $\xi$ and $\eta$ are small fluctuations. Substitution of eq.(21) into the equations of motion (16) and (17) yields the following power series equations 
 of the fluctuation fields $\xi$ and $\eta$, 
\begin{equation}
\label{22}
\hat{l}\left (\begin{array}{c}
\xi\\ \eta \end{array}\right )=\hat{h}\left (\begin{array}{c} \xi \\ \eta \end{array}\right )+\left (\begin{array}{c} G^{\xi}_{2}(\xi ,\eta )\\ G^{\eta}_{2}(\xi
,\eta )\end{array}\right )+\left (\begin{array}{c} G^{\xi}_{3}(\xi ,\eta )\\ G^{\eta}_{3}(\xi
,\eta )\end{array}\right )+\left (\begin{array}{c} G^{\xi}_{4}(\xi ,\eta )\\ G^{\eta}_{4}(\xi
,\eta )\end{array}\right )+\cdots
\end{equation}
where the operators $\hat {l}, \hat {h}$ are defined as
\begin{equation}
\label{23}
\hat {l}=\left (\begin{array}{cc} \frac {d^{2}}{d\tau^{2}} & 0 \\ o & \frac {d^{2}}{d\tau^{2}}\end{array}\right ), \hspace{1cm} \hat {h}=\left (\begin{array}{cc}
 V^{(2)}(\phi_{0}) & 0 \\ o & -8\pi \tilde{G}V(\phi_{0})\end{array}\right )
\end{equation}
and $G_{2}, G_{3},\cdots $ denote terms which contain quadratic,
 cubic and higher powers of the small fluctuations respectively:
\begin{eqnarray*}
 G^{\xi}_{2}&=&-\frac {3}{R_{0}}\left [\dot{\eta}\dot{\xi}
+\eta\ddot{\xi}\right ]+\frac{1}{2}V^{(3)}(\phi_{0})\xi^{2}
+\frac{3}{R_{0}}V^{(2)}(\phi_{0})\xi \eta,\\
 G^{\xi}_{3}&=&-\frac {6}{R^{2}_{0}}\eta \dot{\eta}\dot{\xi}
-\frac {3}{R^{2}_{0}}\eta^{2}\ddot{\xi}+\frac
{1}{3!}V^{(4)}(\phi_{0})\xi^{3}+\frac{3}{2R_{0}}V^{(3)}(\phi_{0})\eta\xi^{2}
+\frac {3}{R_{0}^{2}}V^{(2)}(\phi_{0})\xi\eta^{2},\\
 G^{\xi}_{4}&=&-\frac {3}{R^{3}_{0}}\eta^{2}\dot{\eta}\dot{\xi}
-\frac{1}{R^{3}_{0}}\eta^{3}\ddot{\xi}
+\frac {1}{4!}V^{(5)}(\phi_{0})\xi^{4}+\frac
{1}{2R_{0}}V^{(4)}(\phi_{0})\xi^{3}\eta
+\frac {3}{2R^{2}_{0}}V^{(3)}(\phi_{0})\eta^{2}\xi^{2}
+\frac {1}{R^{3}_{0}}V^{(2)}(\phi_{0})\xi\eta^{3},\\
 G^{\eta}_{2}&=&-\frac {1}{2R_{0}}\dot{\eta}^{2}
-\frac {1}{R_{0}}\eta\ddot{\eta}-2\pi \tilde{G}R_{0}\dot{\xi}^{2}
-2\pi \tilde{G}R_{0}V^{(2)}(\phi_{0})\xi^{2}-\frac {4\pi
\tilde{G}}{R_{0}}V(\phi_{)})\eta^{2},\\
 G^{\eta}_{3}&=&-4\pi \tilde{G}\eta\dot{\xi}^{2}
-4\pi \tilde{G}V^{(2)}(\phi_{0})\eta\xi^{2}-
\frac {4\pi \tilde{G}}{3!}R_{0}V^{(3)}(\phi_{0})\xi^{3},\\
 G^{\eta}_{4}&=&-\frac {2\pi \tilde{G}}{R_{0}}
\eta^{2}\dot{\xi}^{2}-\frac {8\pi 
\tilde{G}}{3!}V^{(3)}(\phi_{0})\eta\xi^{3}
-\frac {2\pi \tilde{G}}{R_{0}}V^{(2)}
(\phi_{0})\xi^{2}\eta^{2}
-\frac {\pi \tilde{G}R_{0}}{3!}V^{(4)}(\phi_{0})\xi^{4}
\end{eqnarray*}
where
 $V^{(n)}(\phi_{0})=\frac{d^{n}V(\phi)}{d\phi^{n}}\vert_{\phi=\phi_{0}}$.
The first--order solution of the fluctuation fields is obvious
 from eq.(22). We have
\begin{equation}
\label{24}
\xi \sim \cos\omega_{0} \tau, \hspace{0.5cm} \eta \sim \cos\omega_{0} \tau, \hspace{0.5cm} \omega^{2}_{0}=\frac {1}{R^{2}_{0}}=-V^{(2)}(\phi_{0})
\end{equation}
where $\omega_{0}$ is the frequency
 of the sphalerons which is simply the frequency
 of small oscillations in the bottoms of the 
inverted potential wells of $U(R)$ and
$V(\phi )$. Substituting the first-order solution into eq.(22)
 we obtain the second--order solution; the higher--order results
 can be obtained by successive
substitutions. After the second substitution we have,
\begin{equation}
\label{25}
\xi =\rho\cos\omega\tau
 +\rho^{2}[g_{1,\xi}+g_{2,\xi}\cos 2\omega \tau ]+\xi_{3},
\end{equation}
\begin{equation}
\label{26}
\eta =\rho\cos\omega\tau +\rho^{2}g_{2,\eta}\cos 2\omega\tau +\eta_{3},
\end{equation}
where
\begin{eqnarray*}
g_{1,\xi} &=& \frac{1}{2\omega^{2}_{0}}
\left [\frac {1}{2}V^{(3)}(\phi_{0})-3\omega^{3}_{0}\right],\\
g_{2,\xi} &=& -\frac{1}{6\omega^{2}_{0}}
\left[3\omega^{3}_{0}+\frac{1}{2}V^{(3)}(\phi_{0})\right],\\
g_{2,\eta} &=& -\frac{1}{3\omega_{0}}
\left[\frac{3}{4}\omega^{2}_{0}+2\pi \tilde{G}(1-V(\phi_{0}))\right].
\end{eqnarray*}
In our case $g_{1,\eta}=0$. Here $\rho$ serves as
 a perturbation parameter. The third--order corrections
 $\xi_{3}, \eta_{3}$ are proportional to $\rho^{3}$.
Substitution of eqs.(25), (26) into the equation of
 motion (22) yields an equation to determine
 $\xi_{3}$ , $\eta_{3}$ and the corresponding frequency $\omega$.
After some tedious algebra we obtain the deviation of
the frequency from $\omega_{0}$ up to order of $\rho^{4}$, i.e. 
\begin{equation}
\label{27}
\omega^{2}-\omega^{2}_{0}=-\rho^{2}\frac{4\pi
\tilde{G}}{3\omega^{2}_{0}}V^{(3)}
(\phi_{0})g_{1,\xi}
-\rho^{4}2\pi \tilde{G}\left[\frac{V^{(4)}(\phi_{0})}{3\omega^{2}_{0}}
+2\omega^{2}_{0}\right]g^{2}_{1,\xi}
\end{equation}
The $\rho^{4}$ term applies in case the $\rho^{2}$ term vanishes.
 The sufficient condition for a transition of the first order to occur
 is $\omega^{2}-\omega^{2}_{0}>0$. In Fig. 2 we show the
phase diagram
taking into account terms up to the order of $\rho^{2}$.

 We
 now analyse some field models in terms of our criterion eq.(27).
 For the well studied $\phi^{4}$ model,
\begin{equation}
\label{28}
V(\phi )=(\phi^{2}-\alpha^{2})^{2}
\end{equation}
we have $\phi_{0}=0$, $\omega_{0}^{2}=-V(\phi_{0})=4\alpha^{2}$,
 $V^{(3)}(\phi_{0})=0$ and $V^{(4)}(\phi_{0})=24$.  Eq.(27) leads to
\begin{equation}
\label{29}
\omega^{2}-\omega^{2}_{0}<0
\end{equation}
The transition is of second order, in agreement
with previous observations
 in the literature\cite{affle,linde}. In recent
 investigations it was pointed out that the transition can be first
order with a steeper well of the potential\cite{kuzne},
\begin{equation}
\label{30} 
V(\phi)=\frac{4+\alpha}{12}-\frac{1}{2}\phi^{2}-\frac{\alpha}{4}\phi^{4}+\frac{1+\alpha}{6}\phi^{6}
\end{equation}
which is a double--well type for $\alpha >0$ (see Fig. 3). The sphaleron is $\phi_{0}=0$ with frequency $\omega_{0}=1$. Since $V^{(3)}(\phi_{0})=0$ the criterion for the
first--order transition is determined by the $\rho^{4}$ term such that
\begin{equation}
\label{31} 
\omega^{2}-\omega^{2}_{0}=-18\pi\rho^{4}[1-\alpha]\tilde{G}
\end{equation}
We thus have either first or second order transitions
 depending on the parameter $\alpha$. When $0<\alpha<1$ 
the transition is of second order, while
for $\alpha>1$ it is of the first order.

\section{Vector matter field}
\label{sec:4}
The winding number transition at finite temperature,
 i. e. the transition between degenerate vacua
 with the vector field model of eq.(5),
 has been investigated recently
using a similar method,
 but in a flat space-time\cite{park1,park2}. It
was
found that in that context 
the transition is always of the first order.
 We now consider the nucleation of the model Universe in
the presence of the same vector field.
 We reexpress the vector field with unit norm  in terms of angular
variables, i.e.
\begin{equation}
\label{32}
{\bf n}=(\sin\theta\cos\varphi, \sin\theta\sin\varphi, \cos\theta)
\end{equation}
The euclidean equations of motion are found from the Lagrangian (15) to be
\begin{equation}
\label{33}
\frac{d}{d\tau}(R^{3}\dot{\theta})
-R^{3}\sin\theta\cos\theta\dot{\varphi}
+\frac{R^{3}}{m\lambda^{2}}\sin\theta=0,
\end{equation}
\begin{equation}
\label{34}
\frac{d}{d\tau}(R\dot{R})-\frac{1+\dot{R}^{2}}{2}
+2\pi GR^{2}\left[m\sum_{a}\dot{n}^{2}_{a}
+\frac{2}{\lambda^{2}}(1+\cos\theta)\right]=0,
\end{equation}
\begin{equation}
\label{35}
\frac{d}{d\tau}(R^{3}\dot{\varphi}\sin^{2}\theta)=0
\end{equation}
where
\begin{eqnarray*}
\dot{n}_{1}&=&\dot{\theta}\cos\theta\cos\varphi
-\dot{\varphi}\sin\theta\sin\varphi,\\
\dot{n}_{2}&=&\dot{\theta}\cos\theta\sin\varphi
+\dot{\varphi}\sin\theta\cos\varphi,\\
\dot{n}_{3}&=&-\dot{\theta}\sin\theta.
\end{eqnarray*}
The sphaleron solution which is obtained from $\dot{\theta}=\ddot{\theta}=\dot{\varphi}=\ddot{\varphi}=\dot{R}=\ddot{R}=0$ is seen to be
\begin{equation}
\label{36}
{\bf n}_{0}=(0, 0, 1), \hspace{1cm} R_{0}=\frac{\lambda}{4\sqrt{\pi G}}
\end{equation}
with $\theta_{0}=0$ and $\varphi_{0}$ an arbitrary constant.
 We again consider the perturbation expansion around
 the sphaleron configurations and set
\begin{equation}
\label{37}
\theta=\theta_{0}+\gamma, \hspace{0.5cm}\varphi=\varphi_{0}+\delta, \hspace{0.5cm}R=R_{0}+\zeta
\end{equation}
A self consistent solution is determined by
\begin{equation}
\label{38}
\hat{l}\left (\begin{array}{c}
\gamma\\ \zeta \end{array}\right )=\hat{h}_{v}\left (\begin{array}{c} \gamma \\ \zeta \end{array}\right )+\left (\begin{array}{c} G^{\gamma}_{2}(\gamma ,\zeta )\\
G^{\zeta}_{2}(\gamma
,\zeta )\end{array}\right )+\left (\begin{array}{c} G^{\gamma}_{3}(\gamma ,\zeta )\\ G^{\zeta}_{3}(\gamma
,\zeta )\end{array}\right )+\left (\begin{array}{c} G^{\gamma}_{4}(\gamma ,\zeta )\\ G^{\zeta}_{4}(\gamma
,\zeta )\end{array}\right )+\cdots
\end{equation}
with $\delta=const.$ where
\begin{equation}
\label{39}
 \hat {h}_{v}=\left (\begin{array}{cc}
 -\frac{1}{m\lambda^{2}} & 0 \\ o & -\frac{16\pi G}{\lambda^{2}}\end{array}\right )
\end{equation}
and
\begin{eqnarray*}
 G^{\gamma}_{2}&=&-\frac {3}{R_{0}}
\left [\dot{\zeta}\dot{\gamma}
+\zeta\ddot{\gamma}+\frac{1}{m\lambda^{2}}\zeta\gamma\right ],\\
 G^{\gamma}_{3}&=&-\frac {6}{R^{2}_{0}}\zeta \dot{\zeta}\dot{\gamma}
-\frac {3}{R^{2}_{0}}\zeta^{2}\ddot{\gamma}+\frac
 {1}{3!m\lambda^{2}}\gamma^{3}
+\gamma\dot{\delta}^{2}-\frac {3}{R_{0}^{2}m\lambda^{2}}\gamma\zeta^{2},\\
 G^{\gamma}_{4}&=&-\frac {3}{R^{3}_{0}}\zeta^{2}\dot{\zeta}\dot{\gamma}
-\frac{1}{R^{3}_{0}}\zeta^{3}\ddot{\gamma}
+\frac{3}{R_{0}}\zeta\dot{\delta}^{2}\gamma+
\frac {1}{2m\lambda^{2}R_{0}}\zeta\gamma^{3}-\frac
{1}{m\lambda^{2}R_{0}^{3}}\zeta^{3}\gamma,\\
 G^{\zeta}_{2}&=&-2\pi GmR_{0}\dot{\gamma}^{2}
-\frac {1}{R_{0}}\zeta\ddot{\zeta}
+\frac{2\pi G}{\lambda^{2}}R_{0}\gamma^{2}-\frac{8\pi
 G}{\lambda^{2}R_{0}}\zeta^{2}
-\frac {1}{2R_{0}}\dot{\zeta}^{2},\\
G^{\zeta}_{3}
&=&-4\pi Gm\zeta\dot{\gamma}^{2}
+\frac{4\pi G}{\lambda^{2}}\zeta\gamma^{2},\\
 G^{\zeta}_{4}&=& -2\pi mGR_{0}\gamma^{2}\dot{\delta}^{2}
-\frac {2\pi Gm}{R_{0}}\zeta^{2}\dot{\gamma}^{2}
-\frac {\pi G}{3!\lambda^{2}}R_{0}\gamma^{4}+
\frac {2\pi G}{\lambda^{2}R_{0}}\zeta^{2}\gamma^{2}.
\end{eqnarray*}
The solution for the fluctuation in first--order approximation is
\begin{equation}
\label{40}
\gamma\sim \cos\omega_{0}\tau, \hspace{1cm} \zeta\sim \cos\omega_{0}\tau
\end{equation}
where the frequency of the sphaleron is found to be
\begin{equation}
\label{41}
\omega_{0}=\frac{4\sqrt{\pi G}}{\lambda}=\frac{1}{R_{0}}, \hspace{1cm} m=\frac{1}{16\pi G}
\end{equation}
The solution for fluctuations up to the third--order approximation is
\begin{equation}
\label{42}
\gamma=\rho\cos\omega\tau-\frac{\rho^{2}}{2}[3\omega_{0}
+\omega_{0}\cos2\omega\tau]+\gamma_{3},
\end{equation}
\begin{equation}
\label{43}
\zeta=\rho\cos\omega\tau-\frac{\rho^{2}}
{6\omega_{0}}(\frac{1}{4}+\omega^{2}_{0})\cos2\omega\tau +\zeta_{3},
\end{equation}
where $\gamma_{3}$, $\zeta_{3}$ and $\omega$
  are determined by substitution of eqs.(42), (43) into the
 equation of motion (38). By doing so the deviation of the 
frequency which we are
interested in is obtained as, 
\begin{equation}
\label{44}
\omega^{2}-\omega^{2}_{0}=\rho^{2}\frac{3\omega^{2}_{0}}{2}(1+3\omega^{2}_{0})>0
\end{equation}
which is always positive. The transition is
therefore of first order, and is, remarkably,
 the same as that of the winding number transition of the
 vector field model in a flat space--time.

\section{Conclusion}
\label{sec:5}
We believe that the present study is the first attempt to
 investigate the nucleation of a RW closed Universe at
 finite temperature with time--dependent matter fields.
Although we consider only the crossover behaviour of the
 nucleation from quantum tunneling to thermal activation,
 this investigation may shed light on the understanding of
the time--evolution of the early Universe. Unlike
 the superminispace model in which only the first
 order transition exists, we find that both first and second order
transitions are possible here, depending on the shape of the
potential of the matter fields.  From another point of view
 the system considered here can be regarded as the barrier
penetration of the field models in the closed RW metric.
 A remarkable observation is that the crossover behaviours,
 i.e. (1)the second order transition for the ordinary
$\phi^{4}$ double--well potential of the scalar field,
 (2) both the first and second for a steeper double--well potential,
but (3) first order only for the $O(3)$ nonlinear
$\sigma$ model, maintain the same relations as those of transitions 
of these field models in a flat space-time. 

\vspace{2cm}
{\bf Acknowledgements} J.--Q.L. and D.K.P. acknowledge support 
by the Deutsche Forschungsgemeinschaft. J.--Q.L. also acknowledges
 partial support by the National Natural Science Foundation of China
under Grant No. 19775033.

\newpage
\begin{center}
{\bf Figure Captions}
\end{center}
\noindent
Fig. 1: Barrier of nucleation and the sphaleron $R_{0}$\\
Fig. 2: Phase diagram with scalar field. I. first order region. II. second order region.\\
Fig. 3: Steeper double--well potentials with $\alpha =0.1$  and  $4$

\end{document}